\documentclass[aps,prb,twocolumn,showpacs]{revtex4}
\usepackage{graphicx}

\begin{document}

\title{Huge negative differential conductance in Au-H$_2$ molecular nanojunctions}

\author{A.~Halbritter, P.~Makk, Sz.~Csonka, and G.~Mih\'aly}
\affiliation{Department of Physics, Budapest University of
Technology and Economics and Condensed Matter Research Group of the Hungarian Academy of Sciences, 1111 Budapest, Budafoki ut 8., Hungary}

\date{\today}

\begin{abstract}
Experimental results showing huge negative differential
conductance in gold-hydrogen molecular nanojunctions are
presented. The results are analyzed in terms of two-level system
(TLS) models: it is shown that a simple TLS model cannot produce
peaklike structures in the differential conductance curves,
whereas an asymmetrically coupled TLS model gives perfect fit to
the data. Our analysis implies that the excitation of a bound
molecule to a large number of energetically similar loosely bound states is
responsible for the peaklike structures. Recent experimental
studies showing related features are discussed within the
framework of our model.
\end{abstract}


\maketitle

\section{Introduction}

The study of molecular nanojunctions built from simple molecules
has attracted wide interest in recent years.\cite{agrait}
Contrary to more complex molecular electronics structures, the
behaviour of some simple molecules bridging atomic-sized metallic
junctions can already be understood in great detail, including the
number of conductance channel analysis with conductance
fluctuation and shot-noise
measurements,\cite{smit,djukicshotnoise} the identification of
various vibrational modes with point-contact spectroscopy
\cite{djukic} and the quantitative predictive power of computer
simulations.\cite{thygesen} The above methods were successfully
used to describe platinum-hydrogen junctions: it was shown that a
molecular hydrogen bridge with a single, perfectly transmitting
channel is formed between the platinum electrodes.\cite{smit}

\begin{figure}
\centering
\includegraphics[width=0.7\columnwidth]{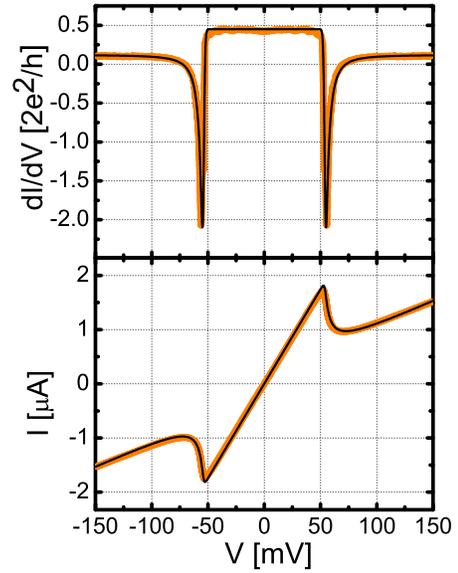}
\caption{\it Huge negative differential conductance (upper panel)
and the corresponding $I-V$ curve (lower panel) in gold-hydrogen
nanojunctions. The solid black lines show the theoretical fits
with the asymmetrically coupled TLS model. The parametrs are:
$E=54$\,meV, $T=5.1$\,K, $\sigma _0=0.45\,$G$_0$, $\sigma
_{\infty}=0.12$\,G$_0$, $N=60$ and $W=0$.} \label{NDR.fig}
\end{figure}

Point-contact spectroscopy turned out to be an especially useful
tool in the study of molecular nanojunctions, as a fingerprint of
the molecular vibrational modes can be given by simply identifying
the small steplike signals in the $dI/dV(V)$ curves of the
junction.\cite{djukic} However, the detection of the small vibrational
signals is difficult for junctions with partially transmitting
conductance channels, where quantum interference (QI) fluctuations may give
an order of a magnitude larger signal.\cite{ludoph} Surprisingly, instead of observing small
steplike vibrational signals upon the background of QI
fluctuations, molecular nanojunctions frequently show peaklike structures
in the differential conductance curves with amplitudes comparable to or even much
larger than the QI
fluctuations.
The peaks can be either positive or negative and their amplitude can
be as small as a few percents, but -- as we demonstrate in this
manuscript -- the peak-height can be several orders of magnitude
larger showing a huge negative differential conductance (NDC)
phenomenon (Fig.~\ref{NDR.fig}).

As a general feature of the phenomenon it can be stated that at
low bias the conductance starts from a constant value, at a
certain threshold voltage the peaklike structure is observed, and
at higher bias the conductance saturates at a constant value
again. The high-bias conductance plateau can be either higher or
lower than the low-bias plateau. In the first case the peak at the
transition energy is positive, whereas in the second case it is
negative. In other words, the junction shows linear $I-V$
characteristics both at low and high biases but with a different
slope, and at the threshold energy a sharp transition occurs
between the two slopes. The switching of the conductance between
two discrete levels implies the description of the phenomenon with
a two-level system (TLS) model, but as we demonstrate in this
paper a simple TLS model only gives steps in the $dI/dV$ at the
excitation energy of the TLS, and it cannot account for sharp
peaks.

The above phenomenon is not a unique feature of a special atomic
configuration, but it frequently occurs in a wide
conductance range and it appears almost in all the studied
molecules and contact materials. In our experiments on
gold-hydrogen molecular junctions\cite{csonka} we have recognized
huge negative differential conductance curves in the conductance
regime of $0.1-0.7$\,G$_0$ showing a characteristic threshold
energy of $30-100$\,meV. We have also observed similar phenomenon
in niobium-hydrogen nanojunctions. In parallel the group of Jan
van Ruitenbeek -- focusing on the conductance range close to the
quantum conductance unit -- has demonstrated the occurrence of
smaller peaklike structures.\cite{thijssen} In their study the
peaks appeared with H$_2$, D$_2$, O$_2$, C$_2$H$_2$, CO, H$_2$O
and benzene as molecules, and Pt, Au, Ag and Ni as contact
electrodes. An scanning tunneling microscopy (STM) study of a hydrogen-covered Cu surface revealed
both sharp gaplike positive peaks and large negative differential
conductance peaks in the tunneling regime.\cite{gupta} The latter
study has also demonstrated a clear telegraph fluctuation between
the two levels on the millisecond timescale. An earlier STM study
demonstrated smaller negative differential conductance peaks due
to the conformational change of pyrrolidine molecule on copper
surface.\cite{gaudioso}

The above results imply the presence of a general physical
phenomenon of molecular nanojunctions resulting in a similar
feature under a wide range of experimental conditions. In
Ref.~\onlinecite{thijssen} it was shown that in certain systems
the peaklike structures are related to the vibrational modes of
the molecular junctions and the observed features were
successfully described in terms of a vibrationally mediated
two-level transition model. This model, however, cannot describe
the large negative differential conductance curves in our studies,
and the large peaklike structures in the STM
experiments.\cite{gupta}

In the following, we present a detailed analysis of the
observations in terms of two-level system models. We demonstrate
the failure of a simple TLS model and the necessary ingredients
for producing peaklike structures instead of simple conductance
steps. We propose a model based on an asymmetrically coupled
two-level system, which can describe all the above observations.
The comparison of the model with experimental data implies a huge
asymmetry in the coupling strength, which can be explained by exciting
a strongly bound molecule to a large number of energetically similar loosely
bound states.
In our opinion, the proposed model is appropriate for describing the appearance
of peaklike structures (or even negative differential conductance)
in the $dI/dV$ curves of molecular nanojunctions under a wide range of experimental
conditions.

\section{The failure of a simple two-level system model}

\begin{figure}
\centering
\includegraphics[width=\columnwidth]{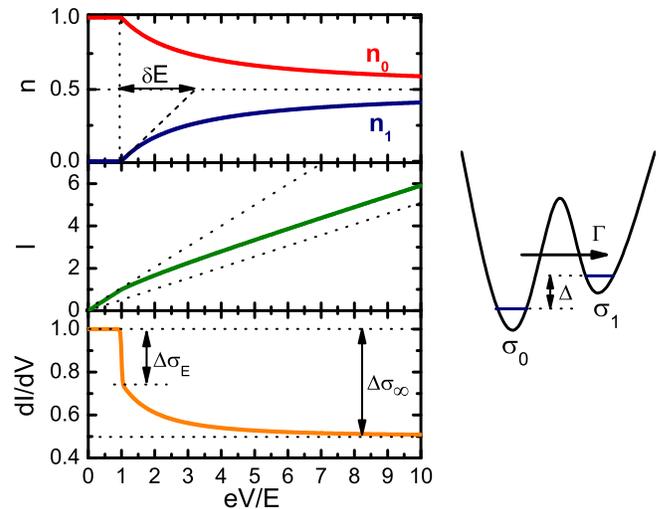}
\caption{\it Results in the zero temperature limit of a
symmetrically coupled TLS model. The upper panel shows the voltage
dependence of the occupation numbers. The characteristic energy
describing the growth of $n_1$ above the excitation energy is
denoted by $\delta E$. The middle panel shows the $I-V$ curve, the
low and high bias slopes ($\sigma_0 \cdot V$, $\sigma_{\infty}
\cdot V$) are illustrated by dotted lines. The lower panel shows
the differential conductance exhibiting a jump of $\Delta\sigma_E$
at the excitation energy and an overall change of
$\Delta\sigma_{\infty}$ between the zero and high bias limits. On
the right side the two-level system model is demonstrated.}
\label{TLSfig1.fig}
\end{figure}

The scattering on two-level systems in mesoscopic point-contacts
has been widely studied after the discovery of point-contact
spectroscopy (see Ref.~\onlinecite{halbritter} and references
therein). In the following, based on the results for a general
point-contact geometry,\cite{halbritter} we shortly give an overview of the
scattering process on a two-level system located near the center
of an atomic-sized nanojunction. The TLS is considered as a double
well potential, where the two states in the two potential wells
have an energy difference $\Delta$ (see Fig.~\ref{TLSfig1.fig}).
The two states are coupled by tunneling across the barrier with a
coupling energy $\Gamma$. The coupling between the wells causes a
hybridization of the two states, resulting in two energy
eigenvalues with a splitting of $E=\sqrt{\Delta^2+\Gamma^2}$. It
is considered that in the lower state of the TLS the contact has a
conductance $\sigma_0$, whereas in the upper state the conductance
is $\sigma_1$. The occupation number of the two states are denoted
by $n_0$ and $n_1=1-n_0$. These are time-averaged occupation
numbers, the system is jumping between the two states showing a
telegraph fluctuation. On a time-scale much longer than that of
the TLS the fluctuation is averaged out, and the $I-V$
characteristic is determined by the voltage dependent occupation
numbers:
\begin{eqnarray}
\label{IV.eq}I(V)&=&( \sigma _0 n_0 + \sigma _1 n_1 ) V, \\
\label{dIdV.eq} \frac{dI}{dV}&=&\sigma_0 + (\sigma _1 - \sigma
_0)\left[n_1+ V \frac{dn _1}{dV} \right].
\end{eqnarray}
The occupation of the upper state can be calculated from the rate
equation:
\begin{equation}
\frac{dn_1}{dt}=P_{0\rightarrow 1}-P_{1\rightarrow 0},
\end{equation}
where the transition probabilities are determined by Fermi's
golden rule:
\begin{eqnarray}
P_{0\rightarrow 1}=\rho_F^2 n_0 \gamma\int\textrm{d}\epsilon
f(\epsilon, eV)(1-f(\epsilon-E, eV)),\label{P01.eq}\\
P_{1\rightarrow 0}=\rho_F^2 n_1 \gamma\int\textrm{d}\epsilon
f(\epsilon, eV)(1-f(\epsilon+E, eV)) \label{P10.eq}.
\end{eqnarray}
Here an electron from an occupied state with energy $\epsilon$
scatters on the TLS to an unoccupied state with energy $\epsilon
\pm E$. The nonequilibrium distribution function of the electrons
is denoted by $f(\epsilon, eV)$, whereas $\rho_F$ stands for the
density of states at the Fermi energy. The transition matrix
element from an initial electron state and the ground state of the
TLS to a final electron state and the excited state of the TLS is
considered as a constant coupling strength:
$\gamma=2\pi/\hbar\cdot|\langle i,0 | H_{e-TLS} | f,1 \rangle|^2$.
Assuming that the TLS is situated at the middle of the contact,
where the half of the electrons is coming from the left and the
half from the right electrode, the nonequilibrium distribution
function can be approximated as:
\begin{equation}
f(\epsilon,eV)=\frac{f_0^L(\epsilon)+f_0^R(\epsilon)}{2},
\end{equation}
where $f_0^L(\epsilon)$ and $f_0^R(\epsilon)=f_0^L(\epsilon-eV)$
are the equilibrium Fermi functions of the left and right
electrodes, for which the chemical potential is shifted by the
applied voltage. By inserting the distribution function to
Eq.~\ref{P01.eq}-\ref{P10.eq} and using the formula:
\begin{equation}
\int f_0 ( \epsilon )(1-f_0(\epsilon -a)) \textrm{d}\epsilon
=\frac {a}{2}(\coth{\frac{a}{2kT}}-1)
\end{equation}
the rate equation can be written as:
\begin{equation}
\frac{dn_1}{dt}=n_0\nu_0-n_1\nu_1,
\end{equation}
where $\nu_0$ and $\nu_1$ are the inverse relaxation times of the
lower and upper state of the TLS,
\begin{eqnarray}
\label{nu0.eq} \lefteqn{\nu_0=\frac{\rho_F^2\gamma}{4}
\left[\frac{eV+E}{2}(\coth{\frac{eV+E}{2kT}}-1) +\right.} \\
\nonumber && +\left.\frac{-eV+E}{2}(\coth{\frac{-eV+E}{2kT}}-1) +
E(\coth{\frac{E}{2kT}}-1)\right] \\
\label{nu1.eq} \lefteqn{\nu_1=\frac{\rho_F^2\gamma}{4}
\left[\frac{eV-E}{2}(\coth{\frac{eV-E}{2kT}}-1) +\right.} \\
\nonumber && +\left.\frac{-eV-E}{2}(\coth{\frac{-eV-E}{2kT}}-1) -
E(\coth{\frac{E}{2kT}}-1) \right].
\end{eqnarray}
The steady state solution for the occupation number of the upper
state is:
\begin{equation}
\label{n1.eq} n_1=1-n_0=\frac{\nu_0}{\nu_0+\nu_1}.
\end{equation}
In the zero temperature limit the inverse relaxation times and the
occupation number take the following simple forms:
\begin{eqnarray}
\nu_0=\frac{\rho_F^2\gamma}{4}\cdot\left\{\begin{array}{ll} 0 & \mbox{for $e|V| < E$} \\
e|V|-E & \mbox{for $e|V| \geq E$} \end{array} \right. \\
\label{nu1T0.eq} \nu_1=\frac{\rho_F^2\gamma}{4}\cdot\left\{\begin{array}{ll} 4E & \mbox{for $e|V| < E$} \\
e|V|+3E & \mbox{for $e|V| \geq E$} \end{array} \right. \\
n_1=\left\{\begin{array}{ll} 0 & \mbox{for $e|V| < E$} \\
\frac{1}{2}- \frac{E}{e|V|+E} & \mbox{for $e|V| \geq E$}
\end{array} \right.
\end{eqnarray}
Fig.~\ref{TLSfig1.fig} shows the evolution of the occupation
numbers, the $I-V$ curve and the differential conductance curve in
the zero temperature limit. At $eV\gg E$ both states are equally
occupied with $n=1/2$, but the transition towards this is very
slow. A characteristic energy scale, $\delta E$ for the variation
of the occupation numbers can be defined by extrapolating the
linear growth of $n_1$ at $eV=E^+$ to the saturation value (see
the upper panel in the figure). Due to the small slope of $n_1(V)$
this characteristic energy scale is larger than the excitation
energy, precisely: $\delta E=2E$. This slow change is reflected by
a smooth variation of the $I-V$ curve. The differential
conductance curve shows a steplike change at $eV=E$, and then
saturates to $\sigma_{\infty}=(\sigma_0+\sigma_1)/2$. The size of
the step is smaller than the overall change of the conductance
($\Delta\sigma_E=\Delta\sigma_{\infty}/2$), thus only a step and
no peak is observed in the $dI/dV$ curve.

In the above considerations the calculation of the current
(Eq.~\ref{IV.eq}) includes only elastic scattering of the
electrons on the two states of the TLS with different
corresponding conductances, and the inelastic scattering processes
are just setting the voltage dependence of the occupation numbers.
However, the inelastic scattering of the electrons in principle
gives direct contribution to the current: above the excitation
energy the backscattering on the contact is enhanced due to the
possibility for inelastic scattering. More precisely the inelastic
current correction can be written as:
\begin{eqnarray}
\delta I^{in}=e\cdot [(-P_{{\rm L_+},0\rightarrow
{\rm L_-},1}+P_{{\rm L_+},0\rightarrow {\rm R_+},1} \\
\nonumber -P_{{\rm R_-},0\rightarrow {\rm L_-},1}+P_{{\rm
R_-},0\rightarrow {\rm R_+},1})+(_{0\leftrightarrow 1})].
\end{eqnarray}
Here L$_+$ and R$_-$ are incoming states to the contact on the
left and right side, while R$_+$ and L$_-$ are outgoing states,
and e.g.\ $P_{{\rm L_+},0\rightarrow {\rm L_-},1}$ is the
probability that an electron from an incoming state on the left
side excites the TLS and scatters to an outgoing state on the left
side. The incoming states have the distribution function of the
corresponding reservoir, while the distribution functions of the
outgoing states are a mixture of the left and right Fermi
functions with the transmission probability of the contact,
$\tau$:
\begin{figure}[h!]
\includegraphics[width=0.6\columnwidth]{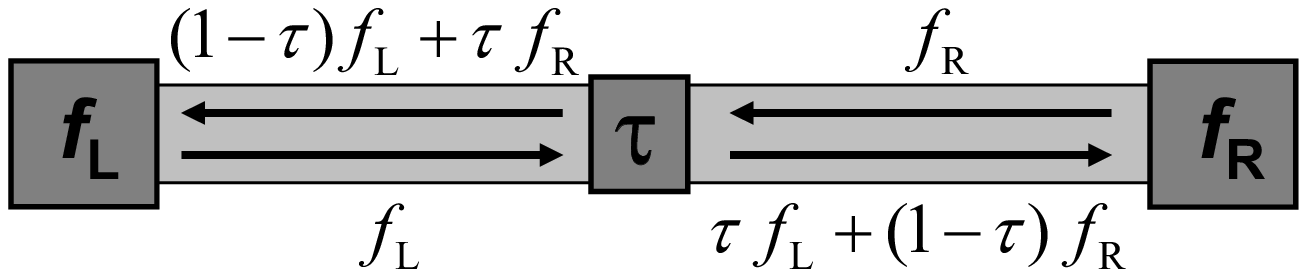}
\end{figure}

\noindent The sign of the different current terms are determined
by the direction of the outgoing states. After evaluating the
energy integrals with the appropriate distribution functions one
obtains:
\begin{equation}
\delta I^{in}=n_0\delta I^{in}_0+n_1\delta I^{in}_1,
\end{equation}
where the first and second term correspond to the excitation and
relaxation of the TLS, and:
\begin{eqnarray}
\lefteqn{\textstyle \delta I^{in}_0=-\frac{e\rho_F^2\gamma(1-2\tau)}{4}\cdot}\\
\nonumber && \textstyle  \left[ \frac{eV+E}{2}(\coth{\frac{eV+E}{2kT}}-1)  - \frac{-eV+E}{2}(\coth{\frac{-eV+E}{2kT}}-1) 
\right]\\
\lefteqn{\textstyle \delta I^{in}_1=-\frac{e\rho_F^2\gamma(1-2\tau)}{4}\cdot}\\
\nonumber && \textstyle \left[\frac{eV-E}{2}(\coth{\frac{eV-E}{2kT}}-1) - \frac{-eV-E}{2}(\coth{\frac{-eV-E}{2kT}}-1) 
\right].
\end{eqnarray}
The inelastic correction also causes a steplike change of the
conductance, which has a magnitude of $\delta G^{in}\approx
e^2\rho_F^2\gamma(1-2\tau)/4$. For a contact with large
transmission the conductance decreases at the excitation energy,
while for a tunnel junction it increases, with a transition
between the two cases at $\tau=1/2$. We note that our simple model
gives the same result for the transition between point-contact
spectroscopy and inelastic electron tunneling spectroscopy at
$\tau=1/2$ as more detailed calculations for vibrational
spectroscopy signals.\cite{paulsson,vega}

The contribution of the inelastic process can easily be estimated.
According to Eq.~\ref{nu1T0.eq} the relaxation time of the TLS is
$\tau_{\rm TLS}\approx (\rho_F^2\gamma E)^{-1}$, thus the
inelastic correction to the conductance is $|\delta
G^{in}|<e^2/(\tau_{\rm TLS}\cdot E)$. It means that a relaxation
time of $1$\,$\mu$sec corresponds to an inelastic correction
smaller than $10^{-7}$\,G$_0$. In other words a correction of
$0.1$\,G$_0$ would correspond to a TLS with a sub-picosecond
relaxation time. These are unphysical numbers, especially if the
telegraph fluctuation can be resolved, thus in the following the
inelastic correction is neglected.

As a conclusion a simple TLS model cannot produce peaklike
structures in the differential conductance curve. Regardless of
the fine details of the model, for any TLS a characteristic energy
$\delta E \gtrsim E$ is required for the transition, whereas for
the observation of sharp peaks in the differential conductance
$\delta E\ll E$ is desired. As an other consequence of a simple
TLS model, the high voltage conductance is limited by
$\sigma_{\infty}>\sigma_0/2$ due to the equal population of the
states at high bias. This contradicts the observation of
Fig.~\ref{NDR.fig}, where $\sigma_{\infty}\approx\sigma_0/5$. For
such large change of the conductance a population inversion,
$n_1(\infty) \gg n_0(\infty)$ is required.

\section{Two-level system model with asymmetric coupling}

Both a population inversion and a sharp transition of the
occupation numbers can be introduced by inserting an asymmetric
coupling constant to the model, that is the coupling of the lower
state of the TLS to the electrons in Eq.~\ref{P01.eq} is
$\gamma_0$, whereas the coupling to the upper state in
Eq.~\ref{P10.eq} is $\gamma_1 \ll \gamma_0$. The asymmetry
parameter is defined as $N=\gamma_0 /\gamma_1$. With this
modification, once the TLS can be excited, it cannot relax back
easily, thus a sharp transition occurs.

\begin{figure}
\centering
\includegraphics[width=\columnwidth]{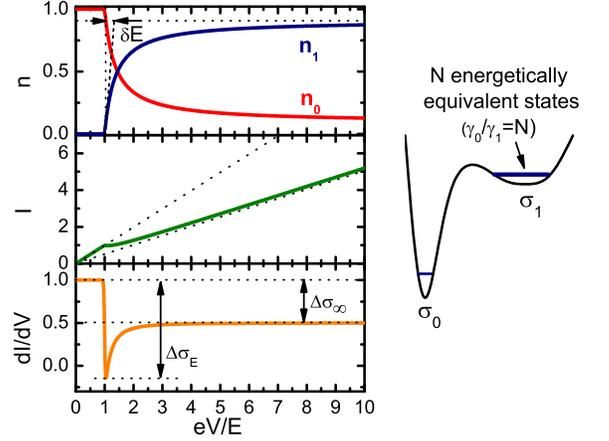}
\caption{\it Results in the zero temperature limit of an
asymmetrically coupled TLS model. The upper panel shows the
voltage dependence of the occupation numbers, the middle panel
shows the $I-V$ curve, and the lower panel shows the differential
conductance. On the right side the two-level system with a
degenerated upper level is demonstrated.} \label{TLSfig2.fig}
\end{figure}

The original definition of the coupling constant does not allow
any asymmetry, as $\gamma_0=2\pi/\hbar\cdot|\langle f,1 |
H_{e-TLS} | i,0 \rangle|^2=2\pi/\hbar\cdot|\langle i,0 | H_{e-TLS}
| f,1 \rangle|^2=\gamma_1$ due to the hermicity of the Hamilton
operator. An asymmetry arises, however, if the phase spaces of the
two levels are different. For instance if the ground state is
well-defined, but the upper state is not a single level, but N
energetically equivalent states (Fig.~\ref{TLSfig2.fig}), then the
effective coupling constant contains a summation for the final
states resulting in $\gamma_0=N\gamma_1$.

With this modification the occupation numbers in the $T=0$ limit
are defined by:
\begin{equation}
n_1=\left\{\begin{array}{ll} 0 & \mbox{for $eV < E$} \\
\frac{\gamma_0(e|V|-E)}{\gamma_0(e|V|-E)+\gamma_1(e|V|+3E)} &
\mbox{for $eV \geq E$}
\end{array} \right. .
\end{equation}
The corresponding occupation numbers, $I-V$ and $dI/dV$ curves are
plotted in Fig.~\ref{TLSfig2.fig}. The population inversion is
obvious with $n_1(\infty)/n_0(\infty)=N$. The characteristic
energy of the variation of the occupation numbers is $\delta
E=4E/(N+1)$. The $I-V$ curve already shows a sharp transition from
the initial to the final slope, and the differential conductance
shows a sharp peak if the asymmetry parameter is high enough. The
appearance of the $dI/dV$ peak is determined more precisely by
calculating the ratio of $\Delta\sigma_E$ and
$\Delta\sigma_{\infty}$:
\begin{eqnarray}
\Delta\sigma_E=\frac{N}{4}(\sigma_1-\sigma_0);&&
\Delta\sigma_{\infty}=\frac{N}{N+1}(\sigma_1-\sigma_0)\ \ \ \ \\
\frac{\Delta\sigma_E}{\Delta\sigma_{\infty}}&=&\frac{N+1}{4}.
\end{eqnarray}
With an asymmetry parameter $N>3$ a peak is observed in the
differential conductance, whereas for smaller asymmetry only a
steplike change arises. For a given difference in the
conductances, $(\sigma_1-\sigma_0)$ both the width and the height
of the conductance peak are determined by the asymmetry parameter,
i.e.\ the peak width and the peak hight are not independent
parameters. Depending on the sign of $(\sigma_1-\sigma_0)$, the
peak can be either positive or negative. The conductance at large
bias is $\sigma_{\infty}=(N\sigma_1+\sigma_0)/(N+1)$, thus it can
be arbitrarily small compared to the zero-bias conductance.

At finite temperature the $dI/dV$ curve is calculated by inserting
Eqs.~\ref{nu0.eq}, \ref{nu1.eq}, \ref{n1.eq} into
Eq.~\ref{dIdV.eq} using asymmetric coupling constants, $\gamma_0,
\gamma_1$. The finite temperature causes a smearing of the curves,
but the transition from steplike to peaklike structure is
similarly observed at $N=3$.

A more realistic generalization of the model can be given by
assuming a distribution of the energy levels at the excited state
with a density of states $\rho_1(E)$, for which the standard
deviation ($W$) is kept much smaller than the mean value ($E_0$).
The asymmetry parameter is defined by the normalization: $N=\int
\rho_1(E){\rm d}E$. With this modification the occupation of the
excited states is energy dependent: $n_1(E)$, but the probability
that any of the excited states is occupied is given by a single
occupation number, $n_1=\int n_1(E){\rm d}E$. The calculation of
the transition probability $P_{0 \rightarrow 1}$ includes the
integration of the excitation rate in Eq.~\ref{nu0.eq} with the
energy distribution:
\begin{equation}
\nu_0(eV,E_0,W)=\int \rho_1(E)\nu_0(eV,E)dE.
\end{equation}
The precise calculation of the relaxation probability ($P_{1
\rightarrow 0}$) would require the knowledge of the energy
dependent occupation number, $n_1(E)$. However, in the narrow
neighborhood of the excitation energy, where the peak is observed
the voltage dependence of the relaxation rate can be neglected
beside the constant value of $\approx 4E$ (see
Eq.~\ref{nu1T0.eq}), thus the original formula for $\nu_1$
(Eq.~\ref{nu1.eq}) is a good approximation for the finite
distribution of the levels as well. Our analysis shows, that the
relaxation rate can even be replaced by a voltage independent
spontaneous relaxation rate without causing significant changes in
the results.

As demonstrated in the following a TLS model with an asymmetry
parameter, $N$, and a narrow width of the upper states, $W$, shows
perfect agreement with the experimental observations.

\section{Experimental results}

In the following we present our experimental results on the
negative differential conductance phenomenon in gold-hydrogen
nanojunctions. The high stability atomic sized Au junctions were
created by low-temperature mechanically controllable break
junction technique. The hydrogen molecules were directed to the
junction from a high purity source through a capillary in the
sample holder. The molecules were dosed by opening a solenoid needle
valve with short voltage pulses, adding a typical amount of $\sim 0.1\,\mu$mol
of hydrogen molecules.

\begin{figure}
\centering
\includegraphics[width=\columnwidth]{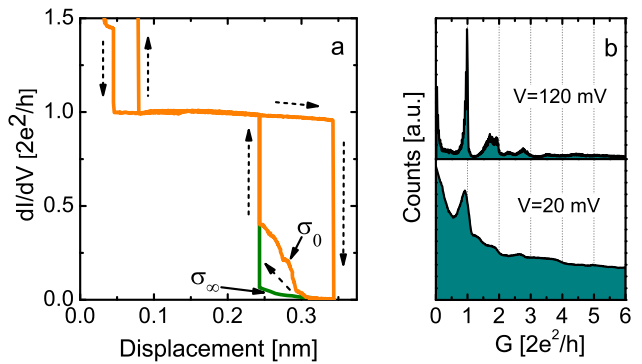}
\caption{\it Panel (a) shows a conductance trace on which NDC
curves were observed when the electrodes were approaching each
other after rupture. In this region the differential conductance
measured at low DC bias ($\sigma_0$, $V_{DC}=0$\,mV) and high bias
($\sigma_{\infty}$, $V_{DC}=120$\,mV) exhibit a large splitting.
The differential conductance was measured with an AC modulation of
$500$\,$\mu$V. Panel (b) shows conductance histograms of Au-H$_2$
junctions measured at a DC bias of $20$\,mV and $120$\,mV,
respectively.} \label{condtrace.fig}
\end{figure}

The appearance of the NDC phenomenon cannot be generally attributed
to definite parts of the conductance traces, but in Au-H$_2$ junctions
the NDC curves were quite frequently observed when the junction
was closed after complete disconnection, as demonstrated by the trace
in Fig.~\ref{condtrace.fig}a. The conductance curve was measured by recording a
large number of $dI/dV(V)$ curves during a single opening -- closing cycle, and
extracting the differential conductance values both at zero bias and at $V_{DC}=120$\,mV.
During the opening of the junctions no difference is observed,
but during the closing of the junction the two curves show large deviation.
The high bias trace resembles the traditional traces of
pure gold junctions: during the approach of the electrodes an
exponential-like growth is observed, but already at a small
conductance value ($<0.1$\,G$_0$) the junctions jumps to a direct
contact with $G=1$\,G$_0$. At low bias voltage the conductance
grows to a much higher value ($\approx 0.5$\,G$_0$) before the
jump to direct contact. The behavior of the conductance trace
in Fig.~\ref{condtrace.fig}a agrees with the general
trend shown by the conductance histograms
(Fig.~\ref{condtrace.fig}b): at low bias a large variety of
configurations is observed at any conductance value, whereas at
high bias the weight in the region $G=0.1-0.8$\,G$_0$ is very
small. Note that only a part of the configurations (a few percent
of all traces) show NDC
features, for other configurations irreversible jumps are observed
in the $I-V$ curve at a certain threshold voltage.
We have not found any evidence that the gold-hydrogen chains reported
in our previous work\cite{csonka} would show NDC phenomenon or any peaklike
structures in the differential conductance curve.

Figure~\ref{NDR.fig} has already presented an example of the huge negative
differential conductance phenomenon. Using the finite temperature results of the
asymmetrically coupled TLS model with a Dirac-delta distribution
of the levels ($W=0$) we have fitted our experimental data in
Fig.~\ref{NDR.fig}. The fitting parameters are the energy, the
temperature and the asymmetry parameter, whereas $\sigma_0$ and
$\sigma_{\infty}$ are directly read from the experimental curve.
As shown by the black solid lines in the figure, the model
provides perfect fit to the data. The striking result of the
fitting is the extremely large asymmetry parameter, $N=60$.

\begin{figure}
\centering
\includegraphics[width=\columnwidth]{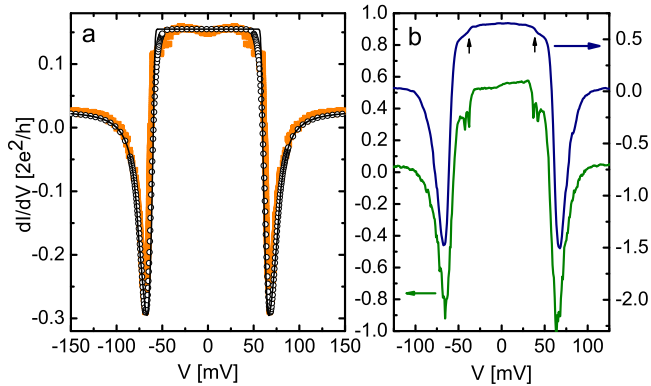}
\caption{\it Panel (a) shows an experimental NDC curve (thick
orange curve in the background) and the fit with a Dirac-delta
distribution (open circles) and a finite width uniform
distribution (thin black line). The fitting parameters are:
$E=66.1$\,meV, $N=42$, $\sigma _0=0.15\,$G$_0$, $\sigma
_{\infty}=0.0022$\,G$_0$; and $T=25 / 4.2$\,K, $W=0 / 4.5$\,meV
for the Dirac-delta and the uniform distribution, respectively.
Panel (b) shows experimental curves exhibiting smaller features
between the NDC peaks.} \label{NDR2.fig}
\end{figure}

Figure~\ref{NDR2.fig}a shows an other example for the NDC
phenomenon. For this curve the fitting
with a Dirac-delta distribution provides a nonrealistic
temperature of $T=25$\,K. By fitting with a finite width uniform
distribution\cite{note1} of the excitation levels the temperature
can be kept at the experimental value ($4.2$\,K) resulting in a
standard deviation of the upper states of $W=4.5$\,meV.
Fig.~\ref{NDR2.fig}b shows further experimental curves exhibiting
smaller features between the negative differential conductance
peaks. In some measurements additional peaks appeared beside the
main NDC peaks, which is demonstrated by the lower curve in the
figure showing two smaller peaks at $E=37.6$ and $42.4$\,meV.
These observations could be explained by introducing further
energy levels with different degeneracy, but such a model would be
too much complicated without aiding the understanding of the
general phenomenon. The upper curve shows a steplike decrease of
$\sim 10\%$ at $E=\pm 39$\,meV resembling the vibrational signal
of molecular junctions, although its amplitude is rather large for
a vibrational spectrum. (The typical conductance change due to
vibrational excitations is $1-3\%$ at the conductance
quantum,\cite{djukic} and it should vanish towards $0.5$\,G$_0$,
where the negative point-contact spectroscopy signal turns to
positive inelastic electron tunneling spectroscopy
signal.\cite{paulsson,vega}) The observed steps could also be
explained by the scattering on a symmetrically coupled TLS, for
which the step-size can be as large as $50\%$ of the zero bias
conductance.

\begin{figure}
\centering
\includegraphics[width=\columnwidth]{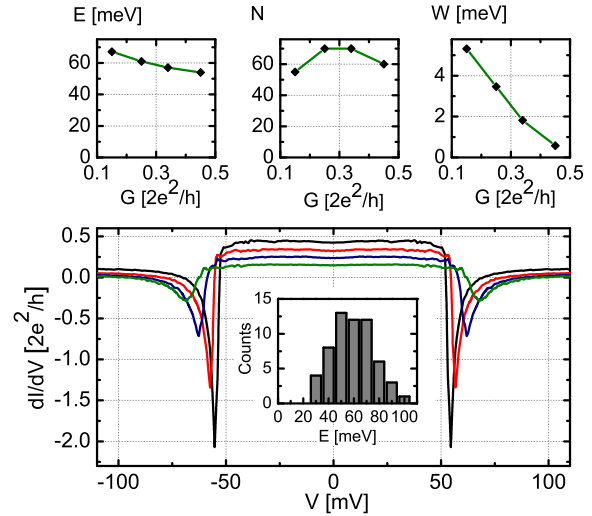}
\caption{\it The lower panel shows the variation on a NDC curve as
a function of electrode separation. The curves were recorded during the
closing of a disconnected junction. The upper panels show the
variation of the fitting parameters as a function of the zero-bias
conductance. The curves with zero bias conductances of $0.25$, $0.35$ and $0.45\,G_0$ correspond to
the pushing of the electrodes by $\approx 0.2\AA$, $\approx 0.4\AA$ and $\approx 0.6\AA$
with respect to the initial curve with zero bias conductance of $0.15\,G_0$.
The inset in the lower panel shows the distribution of the peak positions
for 60 independent NDC curves.} \label{pulling.fig}
\end{figure}

The bottom panel in Fig.~\ref{pulling.fig} shows the variation of the negative
differential conductance curves by changing the electrode
separation. The curves were recorded during the closing of a
disconnected junction, similarly to the trace in
Fig.~\ref{condtrace.fig}a. The upper panels present the change of the fitting
parameters as a function of the zero bias conductance. Both the
excitation energy, $E$ and the width of the excited levels, $W$
decrease with increasing conductance, whereas the asymmetry
parameter, $N$ varies around a constant value.

We have repeated our measurements for a large amount of curves
showing NDC feature. The observed excitation energies have shown a
broad distribution in the range $30-100$\,meV, as demonstrated by the
inset in tho bottom panel of Fig.~\ref{pulling.fig}. Below $30$\,meV
we have not observed any peaks, whereas above $100$\,meV the junctions frequently
become unstable, and the recording of reproducible $I-V$ curves is not possible.
In every case when the electrode separation dependence was studied $E$ was
decreasing by increasing zero-bias conductance showing a shift of the
excitation energy by even a few tens of millivolts.
The standard
deviation of the excited levels was typically in the range
$0.5-10$\,meV, also showing a decreasing tendency with increasing
conductance. The asymmetry parameter was in the range $\approx
40-200$. Note that with these parameters the average spacing of
the excitation energies is smaller than the temperature smearing, thus
the assumption for the continuous distribution of the levels is
correct.

In our experiments the NDC curves on gold-hydrogen junctions have
not shown any telegraph noise within the bandwidth of our setup
($100$\,kHz).

\section{Discussion}

Our analysis of the two-level system models shows that in the case
of two single levels the population of the upper level remains
very small in the narrow neighborhood of the excitation energy,
which inhibits the appearance of sharp peaklike structures in the
differential conductance curve. By introducing a large asymmetry in the
coupling constants, already a small voltage bias above the
excitation energy causes a sudden flip of the occupation numbers,
resulting in peaks or even huge negative differential conductance
in the $dI/dV$ curves.

\begin{figure}
\centering
\includegraphics[width=0.7\columnwidth]{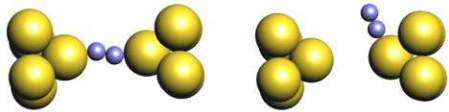}
\caption{\it An illustration for the proposed model: a molecule
strongly bound between the electrodes is excited to a loosely
bound configuration, where different configurations have have
similar binding energy.} \label{TLSillustration.fig}
\end{figure}

The analysis of our experimental curves shows that the negative
differential conductance phenomenon is successfully fitted by the
asymmetrically coupled TLS model, yielding extremely large
asymmetry parameters ($N\approx 40-200$). This result implies,
that the molecular contact has a ground state with a well-defined
molecular configuration, from which it can be excited to a large number of
energetically similar excited states. A trivial explanation would be the desorption
of a bound molecule to the vacuum, however it is hard to imagine,
that after complete desorption a molecule always returns to
exactly the same bound-state. As a more realistic explanation, a
strongly bound molecule is excited into a large number of energetically
similar loosely bound states, from which it can relax to the original
bound state.

A possible illustration is
presented in Fig.~\ref{TLSillustration.fig}. In the ground state
the molecule is strongly bound between the electrodes, whereas the
excited states are loosely bound configurations where e.g. different
angles of the molecule with respect to the contact axis have
similar binding energy, and also the molecule can diffuse to
different sites at the side of the junction. This picture is
consistent with the conductance trace in Fig.~\ref{condtrace.fig}:
at low bias the molecular contact is dominating the conductance,
whereas at high bias the molecule is kicked out to the side of the
contact, and the conductance trace resembles the behavior of pure
gold junctions. It is noted though, that the peaklike structures
are observed with a large diversity of conditions, so in general
Fig.~\ref{TLSillustration.fig} is regarded only as an illustrative picture

The conclusive message of the analysis is, that the
desorption of a strongly bound molecule to a large number of energetically
similar loosely bound states causes peaklike structures or even huge
negative differential conductance in the $dI/dV$ curves. From the fitting of the
experimental curves only the large asymmetry parameter can be deduced,
but the precise microscopic origin of the asymmetry cannot be determined without
detailed microscopic calculations. Possible candidates are the different arrangements
and rotational states of the molecule with respect to the contact, but it
is also very much probable, that the bound molecule is desorbed to
different positions on the side of the contact, and then it can even diffuse away on the
contact surface. In the latter case not necessarily the same molecule
relaxes back to the initial bound state.
All the above processes cause an effective asymmetry in the coupling.

Similar results were obtained on tunnel junctions with higher
resistance in Ref.~\onlinecite{gupta}, where a hydrogen-covered Cu
surface is studied with an STM tip. The authors present a
phenomenological two-state model, which gives perfect fit to the
data, but the microscopic parameters are hidden in the model, and
for the more detailed understanding of the results further
analysis is required. The fitting with our asymmetrically coupled
TLS model shows that the experimental curves in
Ref.~\onlinecite{gupta} correspond to similarly large asymmetry
parameters. In the case of an STM geometry a direct picture can be
associated with the asymmetry: a molecule bound between the
surface and the tip can be excited to several equivalent states on
other sites of the surface, away from the tip. The observation of
the voltage dependent telegraph noise in Ref.~\onlinecite{gupta}
provides a direct measure of the occupation numbers, proving the
population inversion in the system.

\begin{figure}
\centering
\includegraphics[width=0.7\columnwidth]{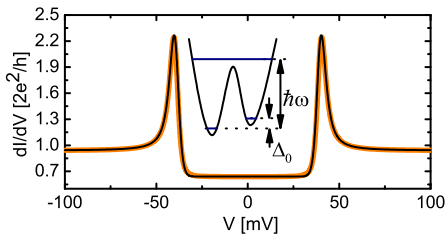}
\caption{\it In the inset the vibrationally mediated TLS model is
illustrated. The thick curve in the background is a fit to one of
the experimental curves in Ref.~\onlinecite{thijssen} using the
vibrationally mediated TLS model (the parameters are:
$\sigma_0=0.63$\,G$_0$, $\sigma_1=1.25$\,G$_0$,
$\hbar\omega=40$\,meV, $\Delta_0=3$\,meV, $T=7$\,K). The thin
black line shows a fit with asymmetrically coupled TLS model with
$E=41$\,meV, $T=4.2$\,K, $N=120$ and $W=2.6$\,meV.}
\label{Ruitenbeek.fig}
\end{figure}

The group of Jan van Ruitenbeek has demonstrated peaklike
structures concentrating on contacts close to the conductance
unit.\cite{thijssen,djukicthesis}
In this conductance range the relative amplitudes of the peaks are
much smaller than our NDC curves in the lower conductance regime, but
the overall shape of the curves is similar. In some cases a direct
transition from a steplike vibrational signal to a peaklike
structure was observed, which implies the coupling of the
phenomenon to molecular vibrations. The coupling to the vibrational modes
was also indicated by the isotope shift of the peak positions. A TLS model illustrated in the
inset of Fig.~\ref{Ruitenbeek.fig} was proposed, in which the two
base levels of the double well potential (with a splitting of
$\Delta_0$) are separated by a wide potential well, which cannot
be crossed. In both wells, however, a vibrational mode of the
molecule can be excited (with $E=\hbar\omega$), and at the
vibrational level the potential barrier can already be crossed,
causing a hybridization of the excited levels between the two
wells. In this model the upper base level is unoccupied for
$eV<\hbar\omega$, but above the excitation energy the transition
between the two wells is possible, and for
$\hbar\omega\gg\Delta_0$ the upper base level suddenly becomes
almost half occupied above the excitation energy. Similarly to the
asymmetrically coupled TLS model, this vibrationally mediated TLS
model produces sharp peaklike structures in the $dI/dV$ due to
the sudden change of the occupation numbers. The authors claim
that the coupling of a vibrational mode to a TLS magnifies the
otherwise tiny vibrational signal, and thus the sharp peaks
directly show the vibrational modes of the junction.

Our analysis shows that the asymmetrically
coupled TLS model proposed by us gives almost identical fits to the
experimental curves in Ref.~\onlinecite{thijssen} as the vibrationally
mediated TLS model, which means that the automatic identification
of the peaklike structures with the vibrational modes of the junction is
only possible if their growth from a
vibrational spectrum is detected, and so the vibrational signal is
anyhow resolved.
In contrast, the negative differential curves demonstrated in our
manuscript cannot be fitted with
the model in Ref.~\onlinecite{thijssen} due to the large ratio of
$\sigma_0/\sigma_{\infty}$ ($\approx 5-50$). The authors use a
symmetric coupling of the two levels, thus
$\sigma_0/\sigma_{\infty}<2$ for moderate voltages
($eV\gtrsim\hbar\omega$), and $\sigma_0/\sigma_{\infty}<3$ for
high bias ($eV\gg\hbar\omega$), where the third level starts to be
populated as well. The model could be generalized by inserting
asymmetric coupling, however in this case the inclusion of the
vibrational level is not necessary any more for fitting the
curves. The broad distribution of the NDC peak positions also indicates
that the phenomenon is not related to vibrational energies.

In our view the vibrationally mediated TLS model
requires unique circumstances: the molecule needs to be really
incorporated in the junction with well-defined vibrational energies.
Furthermore, two similar configurations are required,
which have the same vibrational energy, and for which the transition is only
possible in the excited state. This phenomenon may be general in
certain well-defined molecular junctions, for which the
transition of a vibrational-like signal to a peaklike structure
and the isotope shift of the peaks provides support. In contrast,
the asymmetrically coupled TLS model can be imagined under a much broader range
of experimental conditions, just a bound molecular configuration and a large number of
energetically similar loosely bound (e.g. desorbed) states are required with some
difference in the conductance. The fine details of the underlying physical phenomena
may differ from system to system, however both models provide an illustration of a physical
process that may lead to the appearance of peaklike structures in $dI/dV$ curves of molecular
junctions.

\section{Conclusions}

In conclusion, we have shown that gold-hydrogen nanojunctions
exhibit huge negative differential conductance in the conductance
range of $0.1-0.7$\,G$_0$. The position of the peaks shows a broad
distribution in the energy range $30-100$\,meV. Similar features
were observed in tunnel junctions with higher resistance using an
STM setup,\cite{gupta} whereas in break-junctions with $G\sim
1$\,G$_0$ peaklike structures with smaller amplitude were
detected.\cite{thijssen} These results show that sharp peaklike
structures are generally observed in the differential conductance
curves of molecular nanojunctions under a wide range of
experimental conditions: using different molecules, different
electrode material and different contact sizes. All these
experimental results imply the explanation of the observations in
terms of two-level system models.

We have shown that a simple two-level system model cannot produce
peaklike structures, but a TLS model with asymmetric coupling
successfully fits all the experimental data. Our analysis shows
that the peaklike curves correspond to large asymmetry
parameters, which implies that a molecule from a bound state is
excited to a large number of energetically similar loosely bound
states. This
picture provides a physical phenomenon that can appear under a
wide range of experimental conditions and may explain the frequent
occurrence of peaklike structures in $dI/dV$ curves of various
molecular systems.

Our analysis also shows that a recently proposed vibrationally
mediated two-level system model\cite{thijssen,djukicthesis} may be applicable
for certain well-defined molecular configurations, but the general relation of
the conductance peaks to vibrational modes is not possible without further experimental
evidences.

\section*{ACKNOWLEDGEMENTS}

This work has been supported by the Hungarian research funds OTKA
F049330, TS049881. A.~Halbritter is a grantee of the Bolyai
J\'anos Scholarship.


\begin{references}

\bibitem{agrait}
N. Agrait, A.L. Yeyati, J.M. van Ruitenbeek, Physics Reports {\bf
377}, 81-279 (2003).

\bibitem{smit}
R.H.M. Smit, Y. Noat, C. Untiedt, N.D. Lang, M.C. van Hemert, J.M.
van Ruitenbeek, Nature {\bf 419} 906 (2002).

\bibitem{djukicshotnoise}
D. Djukic and J.M. van Ruitenbeek, Nano Letters, {\bf 6} 789
(2006).

\bibitem{djukic}
D. Djukic, K.S. Thygesen, C. Untiedt, R.H.M. Smit, K.W. Jacobsen,
J.M. van Ruitenbeek, Phys. Rev. B. {\bf 71}, 161402(R) (2005).

\bibitem{thygesen}
K.S. Thygesen, K.W. Jacobsen, Phys. Rev. Lett. {\bf 94} 036807
(2005).

\bibitem{ludoph}
B. Ludoph and J.M. van Ruitenbeek, Phys. Rev. B. {\bf 61} 2273 (2000);
B. Ludoph, M.H. Devoret, D. Esteve, C. Urbina and J.M. van Ruitenbeek, Phys. Rev. Lett. {\bf 82} 1530 (1999).

\bibitem{csonka}
Sz. Csonka, A. Halbritter, and G. Mih\'aly, Phys. Rev. B {\bf 73},
075405 (2006).

\bibitem{thijssen}
W.H.A. Thijssen, D. Djukic, A.F. Otte, R.H. Bremmer, J.M. van
Ruitenbeek, Phys. Rev. Lett. {\bf 97}, 226806 (2006)

\bibitem{gupta}
J.A. Gupta, C.P. Lutz, A.J. Heinrich, D.M. Eigler, Phys. Rev. B.
{\bf 71}, 115416 (2005).

\bibitem{gaudioso}
J. Gaudioso, L. J. Lauhon, and W. Ho Phys. Rev. Lett. {\bf 85},
1918 (2000).

\bibitem{halbritter}
A. Halbritter, L. Borda, A. Zawadowski, Advances in Physics {\bf
53}, 939-1010 (2004).

\bibitem{paulsson}
M. Paulsson, T. Frederiksen, M. Brandbyge, Phys. Rev. B. {\bf 72},
201101(R) (2005).

\bibitem{vega}
L. de la Vega, A. Martin-Rodero, N. Agrait, A.L. Yeyati, Phys.
Rev. B. {\bf 73}, 075428 (2006).

\bibitem{note1}
We have found that the model is not sensitive to the choice of the
distribution, all the distributions with a well-defined width
(e.g. Gaussian) provide similar results. Note, that according to
the general definition $W$ is always the standard deviation and
not the width of the distribution.

\bibitem{djukicthesis}
D. Djukic, PhD thesis, Universiteit Leiden (2006).


\end{references}
\end{document}